# Optical Limiter Based on PT-Symmetry Breaking of Reflectionless Modes

Francesco Riboli[1,2], Rodion Kononchuk[3,4], Federico Tommasi[5], Alice Boschetti[2,5], Suwun Suwunnarat[3], Igor Anisimov[6], Ilya Vitebskiy[6], Diederik S. Wiersma[2,5,7], Stefano Cavalieri[5], Tsampikos Kottos[3], Andrey A. Chabanov[4]

[1] Instituto Nazionale di Ottica – Consiglio Nazionale delle Ricerche (INO-CNR), Sesto Fiorentino, Italy
[2] European Laboratory for Nonlinear Spectroscopy (LENS), Sesto Fiorentino, Italy
[3] Department of Physics, Wesleyan University, Middletown, CT, USA.
[4] Department of Physics and Astronomy, University of Texas at San Antonio, San Antonio, TX, USA.
[5] Dipartimento di Fisica e Astronomia, Università degli Studi di Firenze, Sesto Fiorentino, Italy
[6] Sensors Directorate, Air Force Research Laboratory, Wright-Patterson AFB, OH, USA.
[7] Istituto Nazionale di Ricerca Metrologica (INRiM), Torino, Italy

**Abstract:** The application of parity-time (PT) symmetry in optics, especially PT-symmetry breaking, has attracted considerable attention as a novel approach to controlling light propagation. Here, we study optical limiting by a system of two coupled optical cavities with a PT-symmetric spectrum of reflectionless modes. The optical limiting is related to broken PT symmetry due to light-induced changes in one of the cavities. Our experimental implementation utilizes a three-mirror resonator of alternating layers of cryolite and ZnS with a PT-symmetric spectral degeneracy (also known as the exceptional point of degeneracy) of two reflectionless modes formed at a wavelength of 532 nm. Optical limiting is demonstrated by measurements of single 532-nm 6-ns laser pulses and thermo-optical simulation. At fluences below 10 $mJ/cm^2$, the multilayer has a flat-top passband at 532 nm. At higher fluences, laser heating combined with the thermo-optic effect in ZnS leads to cavity detuning and PT-symmetry breaking of the reflectionless modes. As a result, the entire multilayer structure becomes highly reflective, thereby protecting itself from laser-induced damage. The implemented optical limiter is highly scalable, from near UV to beyond 10 µm. The cavity detuning mechanism can differ at much higher limiting thresholds and include nonlinearity.

Optical limiters (OLs) protect optical systems from damage by intense laser radiation by saturating the transmitted light with increasing irradiation or exposure time[1-6]. A thick layer of optical material with small linear and strong nonlinear absorption can act as a limiter by allowing low-intensity light to pass through and absorbing high-intensity radiation[7-14]. However, this traditional limiting scheme has at least two major problems. First, intense absorption can lead to overheating and damage to the limiter. In other words, the limiting threshold (LT), which determines the level at which a limiter begins to act, may be close to or even the same as the limiter's damage threshold (DT). Second, the required LT varies considerably for different applications and is usually well below the LT provided by available nonlinear optical materials.

To overcome the above problems, a nonlinear optical material can be incorporated into photonic structures, such as photonic bandgap materials[15,16], optical resonators[17,18], and coupled-resonator waveguides[19,20]. In particular, a *reflective* broadband OL based on a planar cavity with nonlinear absorption[21,22] (a GaAs layer sandwiched between two $SiO_2/Si_3N_4$ Bragg mirrors) was implemented and studied using 150-fs laser pulses in the near-infrared region[23]. At low input power, the cavity exhibits resonant transmission through the cavity mode. However, at a sufficiently high intensity, nonlinear absorption of GaAs comes into play, leading first to an increase in absorption and then, ultimately, to suppression of the cavity mode along with resonant transmission. Thus, the entire multilayer structure becomes highly reflective (instead of absorbing) over a broad range of wavelengths, thereby preventing overheating of the OL. However, cavity-based OLs also have drawbacks. First, they can absorb heavily in the transition from low-intensity transmission to high-intensity reflection, especially when compared to a stand-alone nonlinear layer. Secondly, their low-intensity transmission is narrowband due to their resonant nature, which puts a limit on the maximum bandwidth of the optical system to be protected. In addition, due to the resonant nature of the low-intensity transmission, the cavity-based OLs require optical materials with purely nonlinear absorption since even tiny linear absorption will be significantly enhanced at the resonance. Such materials exist but can only work in specific frequency ranges (for example, GaAs in the near-infrared).

Recently, the focus of OL research has shifted to photonic structures with exceptional points of degeneracy (EPD)[24]. Exceptional points are spectral singularities in the parameter space of non-Hermitian operators at which both the eigenvalues and their corresponding eigenvectors coalesce[25]. One method for implementing EPD in the spectrum of an operator is to impose parity-time (PT) symmetry[26-28]. A PT-symmetric operator may undergo a phase transition to a spontaneously broken symmetry phase due to a change in parameters; in this case, the point of symmetry breaking is the EPD. The appearance of EPD in photonic systems endows them with unusual optical properties, which arise mainly due to the anomalous parameter dependence of their optical responses at the EPD (for recent reviews, see refs.[29,30]). The EPD OL proposed and numerically tested in ref.[24] utilized two coupled cavities of different Q factors and material with the Kerr nonlinearity. The EPD of *resonant* modes of the photonic structure was achieved by tuning the differential optical loss between the cavities[31,32]. When the input intensity exceeded a certain threshold level, the nonlinear detuning between the cavities caused a PT-symmetry breaking accompanied by a sharp transition from resonant transmission to broadband reflection. Thus, the advantage of the EPD OL lies in fast limiting action, which shortens the limiter activation time and reduces photon absorption during the transition to the reflective state. However, the limiting scheme of ref.[24] does not solve the problem of narrowband low-intensity transmission. In addition, the differential optical cavity loss may not be practical or desirable for parameter tuning.

Here we take a decisive step towards realistic fast broadband OLs by introducing a limiting scheme based on PT-symmetry breaking of the so-called *reflectionless* modes (RMs). RMs are continuous waves that excite the photonic structure without back reflection[33,34]. Concerning OLs, RMs have several valuable properties: first, in the case of small absorption, the absence of back reflection leads to high transmission; secondly, at the RM EPD, the transmission lineshape is flattened, resulting in a passband rather than a resonance; thirdly, the RM EPD can be realized by a relatively simple geometric tuning of the photonic structure. The introduced OL has the photonic structure of a three-mirror resonator composed of alternating layers of cryolite ($Na_3AlF_6$) and zinc sulfide (ZnS) (Fig. 1a,d). The RM EPD is implemented by tuning the resonator's coupled cavities and cryolite/ZnS Bragg mirrors. On the other hand, cavity detuning due to a change in the refractive index of one of the cavities or the angle of incidence leads to a violation of the PT symmetry of RMs and, consequently, to a high reflectance of the multilayer. Each way of capping the optical throughput of the multilayer provides optical functionality — in our case, optical limiting or directional transmission. The optical limiting is demonstrated by measurements of single laser pulses with a wavelength of 532 nm and a duration of 6 ns. At fluences below 10 mJ/cm$^2$, the multilayer exhibits a high transmittance of ~0.8 in a passband associated with the RM EPD. However, at higher fluences, laser heating in combination with the thermo-optic effect in ZnS leads to cavity detuning, PT-symmetry breaking of RMs, and optical limiting. Our experimental results are supported by multiphysics simulations that give insight into the dynamics of optical limiting of the multilayer. The thermally induced cavity detuning does not require very high fluences or temperatures, which provides fast limiting action (on the ns time scale) and excellent protection against overheating. At much higher fluences or for shorter pulses, cavity detuning is self-induced due to the Kerr effect. The introduced optical limiting scheme is highly scalable and can operate in any spectral range from near UV to more than 10 μm.

## Results

### OL design and modeling

The OL studied in this work is schematically shown in Fig. 1a,d. It consists of two (defect) layers, $(ZnS)^6$ and $(Na_3AlF_6)^6$, playing the role of resonant cavities, and three cryolite/ZnS Bragg mirrors. The OL can be modeled as a system of two resonant modes $\omega_1$ and $\omega_2$, coupled to each other at rate $\kappa$ and to respective transmission lines at rates $\gamma_1$ and $\gamma_2$. The $2 \times 2$ scattering matrix $\boldsymbol{S}$ of this setting was evaluated using coupled mode theory (Supplementary Section 1).

For a continuous wave incident from the left, the reflection and transmission coefficients are defined as $r(\omega) \equiv \boldsymbol{S}_{11}(\omega)$ and $t(\omega) \equiv \boldsymbol{S}_{21}(\omega)$, respectively. The (complex) frequencies $\omega_{RZ}^{(1,2)}$, at which $r(\omega_{RZ}) = 0$, correspond to specific solutions of the scattering problem for which there is no back reflection into the input channel (i.e., R-zeros). The $\omega_{RZ}^{(1,2)}$ are determined from the eigenvalue problem for the auxiliary wave operator $\boldsymbol{A}$,

$$\det(\omega_{RZ}\mathbf{1} - \boldsymbol{A}) = 0, \quad \boldsymbol{A} = \begin{pmatrix} \omega_1 & \kappa \\ \kappa & \omega_2 \end{pmatrix} + i\begin{pmatrix} \gamma_1 & 0 \\ 0 & -\gamma_2 \end{pmatrix}, \tag{1}$$

where $\mathbf{1}$ is the $2 \times 2$ identity matrix. When $\omega_{RZ}^{(1,2)}$ are real-valued, the corresponding scattering solutions are steady-state excitations called RMs[33,34], $\omega_{RM}^{(1,2)} = \omega_{RZ}^{(1.2)} \in \mathbb{R}$. In the case of a passive system (no loss or gain), the RMs have unity transmittance, $T\left(\omega_{RM}^{(1,2)}\right) \equiv \left|t\left(\omega_{RM}^{(1,2)}\right)\right|^2 = 1$ (solid green lines in Fig. 1b). Direct diagonalization of $\boldsymbol{A}$ shows that $\omega_{RZ}^{(1,2)}$ are real-valued provided

$$\omega_1 = \omega_2 = \omega_0, \quad \gamma_1 = \gamma_2 = \gamma, \quad \text{and} \quad \kappa \geq \gamma. \tag{2}$$

Here the first two equations are PT-symmetry conditions for $\boldsymbol{A}$, and the last condition indicates a *spontaneous* symmetry breaking point, $\kappa = \gamma$, where the eigenvalues and their corresponding eigenvectors coalesce, forming a second-order EPD at $\omega_{EPD} = \omega_0$ (open circle in Fig. 1b). An essential feature of the RM EPD is a quartically flat transmission lineshape around $\omega_{EPD} = \omega_0$ (Fig. 1c),

$$T_{\kappa=\gamma}(\omega) = \frac{4\gamma^4}{4\gamma^4 + (\omega - \omega_{EPD})^4}\,. \tag{3}$$

Such lineshape provides a near-unity passband, $T_{\kappa=\gamma}(\omega) \approx 1 - 4(\Delta\omega/2\gamma)^4$, where $\Delta\omega = \omega - \omega_{EPD}$, for the frequency detuning $|\Delta\omega| \leq \sqrt{2}\gamma$, which is balanced by a relatively sharp (faster than Lorentzian) transmittance drop for $|\Delta\omega| > \sqrt{2}\gamma$.

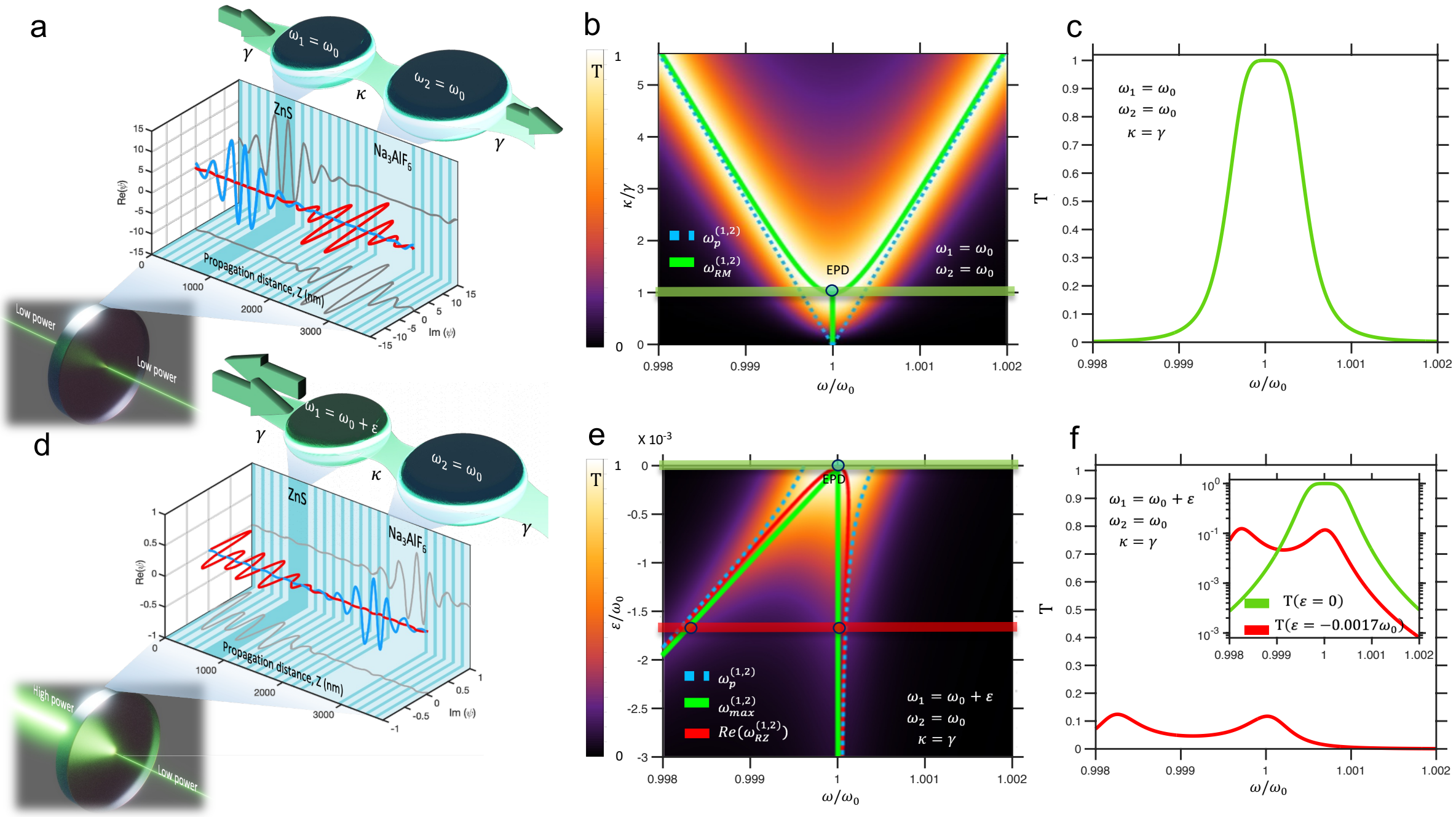

**Fig. 1: OL based on PT-symmetry breaking of RMs. a**, Artistic view (bottom), schematic (middle), and coupled-mode-theory (CMT) equivalent system (top) of the three-mirror resonator composed of cryolite ($Na_3AlF_6$) and zinc sulfide (ZnS) layers. The CMT model involves two resonant modes $\omega_1 = \omega_2 = \omega_0$, coupled to each other at rate $\kappa$ and to respective transmission lines at rates $\gamma_1 = \gamma_2 = \gamma$. **b,** (colormap) Transmittance $T(\omega/\omega_0)$ in the parameter space of $\kappa/\gamma$. $\kappa \geq \gamma$ corresponds to the exact PT-symmetric phase, where the RMs are formed. At frequencies $\omega_{RM}^{(1,2)}$, $T = 1$ (solid green lines). $\kappa < \gamma$ is associated with the spontaneously broken PT-symmetry phase, in which the spectrum consists of the complex-conjugate R-zeros $\omega_{RZ}^{(1,2)}$ (vertical green line). At the spontaneous PT-symmetry breaking point, $\kappa = \gamma$, both the frequencies $\omega_{RM}^{(1,2)}$ and the corresponding RMs $\psi_{RM}^{(1,2)}$ coalesce, forming an EPD at $\omega_{EPD} = \omega_0$ (open circle). The corresponding $\mathcal{Re}(\psi_{EPD})$ and $\text{Im}(\psi_{EPD})$ are shown in (a) by the blue and red lines, respectively. The horizontal green line marks $T_{\kappa=\gamma}(\omega/\omega_0)$, which is plotted in (c). The dashed blue lines indicate the resonance frequencies $\omega_p^{(1,2)}$ (the poles of the $\boldsymbol{S}$-matrix). **c,** Flat-top, near-unity transmittance $T_{\kappa=\gamma}(\omega/\omega_0)$ of Eq. (3). **d,** Same as (a) but in the case of explicit PT-symmetry breaking due to the mode detuning $\omega_1 = \omega_0 + \varepsilon$, when RMs no longer exist. **e,** (colormap) Transmittance $T_{\kappa=\gamma}(\omega/\omega_0)$ in the parameter space of $\varepsilon/\omega_0$. The dashed blue lines indicate the resonance frequencies $\omega_p^{(1,2)}$. The solid green lines denote the peak transmittance frequencies $\omega_{max}^{(1,2)}$. The solid red lines indicate the real part of the R-zeros, $\mathcal{Re}\left\{\omega_{RZ}^{(1,2)}\right\}$. The horizontal red line marks $T_{\kappa=\gamma}(\omega/\omega_0)$ for $\varepsilon/\omega_0 = -0.0017$, which is plotted in (f). **f,** Transmittance $T_{\kappa=\gamma}(\omega/\omega_0)$ of Eq. (4) for $\varepsilon/\omega_0 = -0.0017$. The inset shows a semi-log plot of $T_{\kappa=\gamma}(\omega/\omega_0)$ of (c) and (f).

When one of the resonant modes is detuned, i.e., $\omega_1 \rightarrow \omega_0 + \varepsilon$, the PT symmetry of $\boldsymbol{A}$ is *explicitly* broken. In this domain, RMs no longer exist. Instead, one can consider real-valued resonance frequencies $\omega_{max}^{(1,2)}$, for which the transmittance reaches maxima (solid green lines in Fig. 1e). Specifically, in the proximity of the RM EPD, i.e., when $\kappa \approx \gamma$, the mode detuning results in the transmittance

$$T_{\kappa=\gamma}(\omega) = \frac{4\gamma^4}{4\gamma^4+\gamma^2\varepsilon^2+(\omega-\omega_0)^2(\omega-\omega_0-\varepsilon)^2} \tag{4}$$

with two Lorentzian peaks at $\omega_{max}^{(1)} = \omega_0$ and $\omega_{max}^{(2)} = \omega_0 + \varepsilon$. The Lorentzian peak value $T_{\kappa=\gamma}\left(\omega_{max}^{(1,2)}\right) = \frac{4\gamma^2}{4\gamma^2+\varepsilon^2}$ manifests a reduction in $T$ due to the mode detuning $\varepsilon$. In Fig. 1f, the transmittance $T_{\kappa=\gamma}(\omega)$ of Eq. (4) for $\varepsilon/\omega_0 = -1.7 \times 10^{-3}$ is plotted with a solid red line. The inset shows the comparison of $T_{\kappa=\gamma}(\omega)$ of Eqs. (3) and (4) on a semi-log plot.

One way to detune the coupled cavities of the cryolite/ZnS multilayer is to change the refractive index of the $(ZnS)^6$ or $(Na_3AlF_6)^6$ layer (Fig. 1d). The incident light can self-induce such a change due to nonlinearity or a thermo-optic effect. In the following, we focus on the latter. Another way to break the PT symmetry and lift the RM EPD is to change the angle of incidence. At oblique incidence, the resonant wavelength in each single-mode cavity is “blue-shifted”[35]; however, the shift value in each cavity differs, resulting in RM EPD lifting and corresponding directional transmission.

## Experimental results

Our optical studies are carried out on a sample fabricated according to the multilayer design of Fig. 1 and the RM EPD conditions of Eq. (2) at the design wavelength 532 nm. Omega Optical LLC performed the deposition of the cryolite/ZnS multilayer. The multilayer was deposited on a 25.4-mm diameter, 0.5-mm thick Borofloat glass substrate and laminated to a glass coverslip with an index-matching epoxy (EPO-TEK® 301) to resist moisture and scratches (Supplementary Section 2).

The multilayer was characterized by oblique-angle transmittance measurements using the experimental setup shown schematically in Fig. 2a (Methods section).

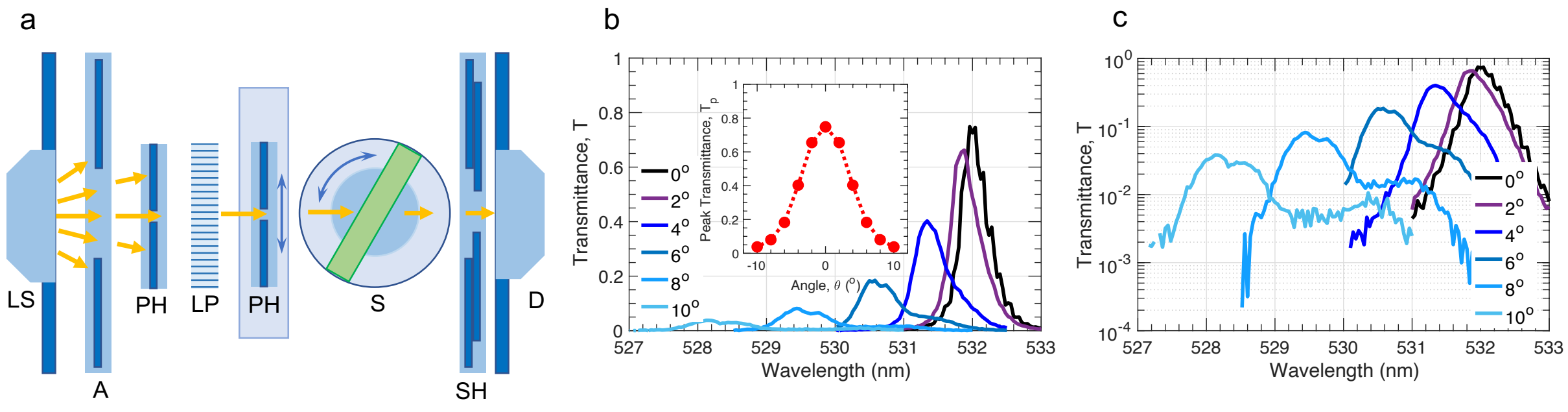

**Fig. 2: Oblique-angle transmittance measurements of the cryolite/ZnS multilayer. a,** Schematic depiction of the experiment setup showing light source (LS), wide aperture (A), pinholes (PH), linear polarizer (LP), sample (S), shutter (SH), and detector (D). **b,** Transmittance spectra for *s*-polarized light as a function of the angle of incidence of the input beam. Inset: angular transmission cone. **c,** Semi-logarithmic plot of the transmittance spectra.

The transmittance spectra measured for *s*-polarized light at an angle of incidence from 0° to 10° are demonstrated in Fig. 2b. At normal incidence, the peak transmittance is 0.75 at 532 nm with a Full-Width Half Maximum (FWHM) of 0.4 nm. At an incidence angle of 10°, peak transmittance drops to less than 0.04, exhibiting highly directional transmission not found in conventional narrow-passband filters. The angular transmission cone of the multilayer is ∼5° (inset in Fig. 2b). Such directional transmission can be used in optical isolators to provide isolation at oblique incidence, lens-free fluorescent imaging[36], and in achieving collimated LED sources[37]. Cavity detuning due to a differential blue-shift between the cavities, leading to PT-symmetry breaking of RMs, is manifested in the increasing splitting of the transmittance peak with the oblique incidence angle. (Fig. 2c). There was no difference (within experimental error) between measurements using $s$- and $p$-polarized light.

Optical limiting was studied using the experimental apparatus shown in the schematics in Fig. 3a (Methods section). In Fig. 3b, the pulse energy transmittance $T$ (blue points) and reflectance $R$ (red points) are plotted versus the incident peak intensity $I_p$ and fluence $F$. At $F < 10$ mJ/cm$^2$, $T$ is at 0.8, and $R$ is slightly lower than 0.2, regardless of $F$, indicating small absorptance in the sample and that the system is in the linear regime (blue-shaded region). Then, the optical limiting becomes effective at $F \geq 10$ mJ/cm$^2$. $T$ decreases and $R$ increases with increasing fluence, reaching $T \approx 0.05$ and $R \approx 0.9$ at $F = 0.7$ J/cm$^2$. After each increase in the energy of the incident pulse, we monitor the state of the sample by performing a recovery-check measurement at the lowest fluence $F = 25$ μJ/cm$^2$. The "recovery" transmittance $T_r$ and reflectance $R_r$, plotted along with the corresponding high-fluence measurement data, are shown by light-blue and light-red points, respectively. The horizontal parts of the $T_r$ and $R_r$ plots correspond to the full recovery of the multilayer. As seen from Fig. 3b, the limiting effect is reversible up to 0.1 J/cm$^2$, above which the multilayer becomes increasingly opaque after exposure to the increased pulse energies (pink-shaded region). This transparency loss of the multilayer, however, is due to laser-induced damage to the adhesive layer[38] and not to the cryolite or ZnS layers, which have much higher DT[39,40].

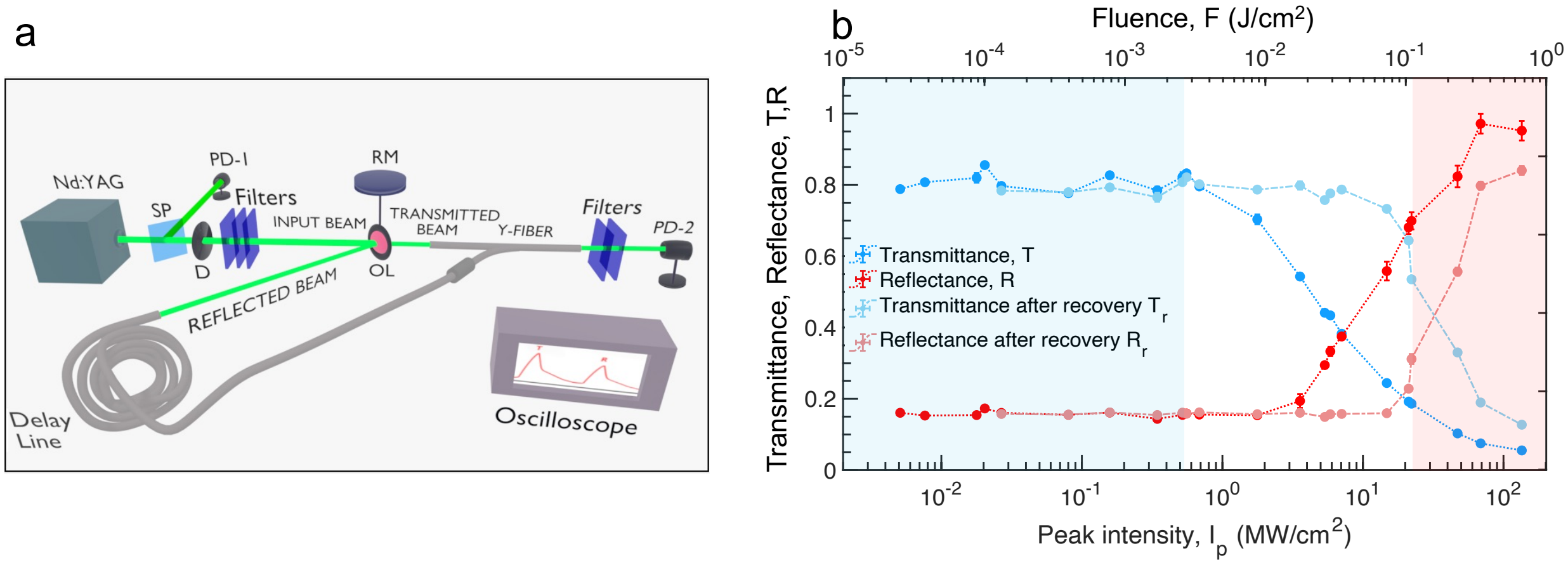

**Fig. 3: Laser-pulse optical limiting measurements. a,** Schematic depiction of the optical limiting experiment showing laser pulse source (Nd:YAG), semi-transparent plate (SP), diaphragm (D), rotary translation mount (RM), sample (OL), photodiode to measure input energy (PD-1), photodiode to measure the transmitted and reflected energies (PD-2). **b,** 532-nm 6-ns laser pulse transmittance $T$ (blue points) and reflectance $R$ (red points) of the cryolite/ZnS multilayer as a function of the incident peak intensity $I_p$ and fluence $F$. Also shown are the "recovery" transmittance $T_r$ (light-blue points) and reflectance $R_r$ (light-red points) measured at the lowest fluence 25 μJ/cm$^2$ to monitor the state of the multilayer after each high-fluence measurement. The horizontal parts of the $T_r$ and $R_r$ plots correspond to the full recovery of the multilayer. The blue-shaded, clear, and pink-shaded regions indicate the linear, optical limiting, and laser-damage regimes, respectively.

The thermo-optic effect is responsible for the limiting action of the multilayer. Indeed, cryolite/ZnS thin-film optical filters exhibit a substantial spectral shift with temperature (up to 37

ppm/°C)[41], which essentially depends on the thermo-optic coefficients of the filter materials and the type of the filter substrate[42]. In the case of the cryolite/ZnS multilayer of Fig. 1a, however, the thermo-optic effect leads to a complete loss of the passband rather than its spectral shift. We note that the resonant cavities of the multilayer can also be detuned due to the optical Kerr effect. However, in our measurements, the fluence is not high enough to cause nonlinearity in cryolite or ZnS[43].

## Thermo-optical simulations

To gain a deeper understanding of optical limiting of the cryolite/ZnS multilayer, we performed thermo-optical simulations of a 5-ns Gaussian pulse in the setting of Fig. 1a (Methods section). In Fig. 4a, the computed transmittance $T$ and reflectance $R$ indicate the limiting effect in the same fluence range as the measured $T$ and $R$ in Fig. 3b. This allows us to extend our analysis to $F = 5$ J/cm$^2$ without ramifications of sample lamination. In Figs. 4b and 4c, we show temperature profiles versus time across the multilayer for the incident pulses of $F = 0.5$ mJ/cm$^2$ (below the LT) and 5 J/cm$^2$ (above the LT), respectively. Below the LT, the temperature profile does not reveal significant laser heating, whereas, above the LT, the temperature rises to $\sim 200$°C in $(ZnS)^6$. This temperature is well below the melting point of ZnS but high enough to cause a frequency detuning between the cavities, leading to optical limiting. Notably, this temperature is confined, on the 10-ns time scale, only to $(ZnS)^6$ since the RMs are exponentially localized in the cavities (Fig. 1a). Thus, the laser heating does not affect the cryolite/ZnS Bragg mirrors. Note that, for simplicity, we do not consider laser heating of the cryolite layers due to their low thermo-optic coefficient.

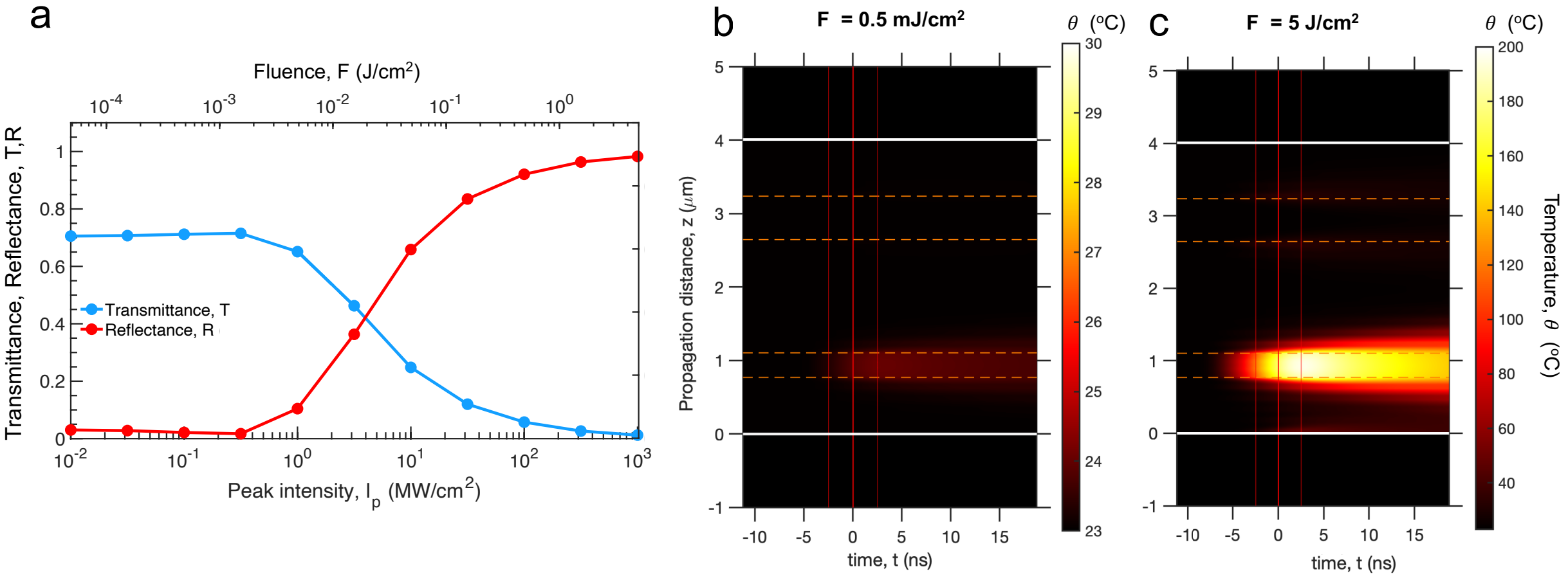

**Fig. 4: Thermo-optical simulations. a,** 532-nm 5-ns laser pulse transmittance $T$ (blue points) and reflectance $R$ (red points) of the cryolite/ZnS multilayer of Fig. 1 as a function of the incident peak intensity $I_p$ and fluence $F$. **b-c,** Temporal temperature profiles across the multilayer for incident pulses of $F = 0.5$ mJ/cm$^2$ (b) and 5 J/cm$^2$ (c). The upper and lower horizontal dashed lines indicate $(Na_3AlF_6)^6$ and $(ZnS)^6$, respectively.

Figures 5a and 5b show the time-resolved transmittance $T(t)$ and reflectance $R(t)$ at the back and front faces of the multilayer, respectively, for several fluence levels. A dashed line denotes the incident pulse whose peak intensity $I_p$ normalizes each intensity profile; here, $I_p$ crosses the front face when $t = 0$. For $F \geq 5$ mJ/cm$^2$, optical limiting starts at the leading edge of the incident pulse (Fig. 5a, negative times) without a significant "leakage spike" (i.e., short-term output power peak). This contrasts with existing thermally activated limiting schemes[44-46], in which a prolonged activation time was considered a significant drawback. Figure 5c shows the absorptance $A(t) = 1 - T(t) - R(t)$ versus time, which drops steadily with increasing fluence, preventing the OL from overheating. Lastly, Figure 5d shows the time evolution of the temperature $\theta(t)$ in the middle of $(ZnS)^6$, following the pulsed excitation. After a sharp temperature rise on the ns time scale, $(ZnS)^6$ cools down in a two-stage process. First, $\theta(t)$ drops to a plateau-like behavior due to heat conductance from $(ZnS)^6$ to the rest of the multilayer. This takes a few microseconds, and the plateau temperature value depends on $F$. Then, the multilayer cools down to the ambient temperature $\theta_0$ due to heat convection at the front and back faces of the multilayer on the time scale dependent on $F$ and $\theta_0$.

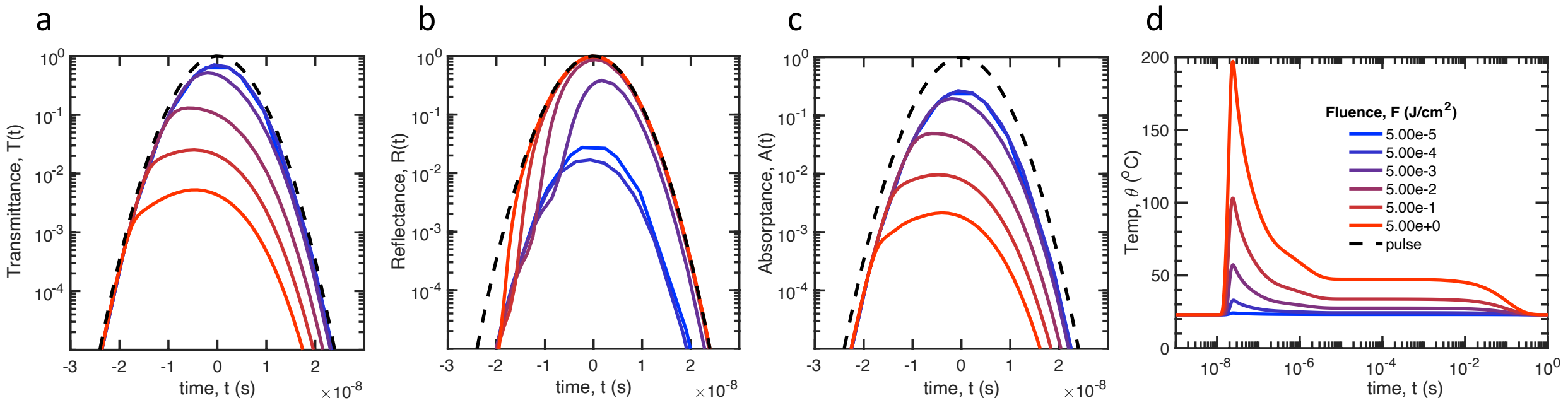

**Fig. 5: Simulated temporal response of the cryolite/ZnS multilayer under pulsed excitation. a-d,** Time evolution of the transmitted intensity $T(t)$ (a), reflected intensity $R(t)$ (b), absorbed intensity, $A(t) = 1 - T(t) - R(t)$ (c), and temperature $\theta(t)$ (d), following a 5-ns laser pulse excitation for a series of fluences (see figure legend). $T(t)$ and $R(t)$ are measured at the back and front faces of the multilayer, respectively, while $\theta(t)$ is determined in the middle of $(ZnS)^6$. The intensity profiles are normalized by the respective incident peak intensities. The black dashed lines are the normalized intensity profile of the incident pulses.

## Discussion

In summary, we have developed an OL design based on PT-symmetry breaking of RMs in a system of two coupled optical cavities. In the PT-symmetric phase, RMs are continuous waves that excite the system without back reflection. However, at high fluences, the PT symmetry is broken due to the laser-induced change in the refractive index of one of the cavities, which suppresses the RMs. Thus, the optical system becomes highly reflective (rather than absorbing) over a broad range of wavelengths, dramatically enhancing the dynamic range of the limiter by preventing it from overheating.

The introduced optical limiting scheme has several critical advantages over the existing OL designs. First, the low-intensity transmittance has a broad flat-top passband at the spectral point of PT-symmetry breaking. In fact, there is no hard limit on the maximum bandwidth of the low-intensity transmittance as long as we can use a PT-symmetric array of multiple coupled cavities supporting RMs. This significantly alleviates the fundamental problem of narrow bandwidth associated with the existing designs of single-cavity resonant limiters. Second, the proposed scheme can use both low-loss optical materials with the Kerr nonlinearity or materials displaying the thermo-optic effect. In the latter case, a much lower limiting threshold and fast limiting action (on the ns time scale) can be achieved at the expense of losses associated with heating the thermo-optic material. In our experiment, the optical limiting associated with the PT-symmetry breaking is caused by the thermo-optic effect in ZnS. Third, the OL based on the thermo-optic effect is highly scalable and can operate in any spectral range from near UV to more than 10 μm. Furthermore, since ZnS and cryolite are a classic film pair in optical bandpass filters, cryolite/ZnS multilayers may combine optical limiting and filtering on the same platform in a wide spectral range.

The optical limiting was demonstrated by normal-incidence measurements of single laser pulses with a wavelength of 532 nm and a duration of 6 ns. However, at oblique incidence, one must consider the differential blue-shift between the cavities in addition to the cavity detuning due to the thermo-optical effect. The two effects can complement or oppose each other. In the latter case, and in our experimental implementation, they may cancel out exactly, leading to a spectral shift of the transmittance passband rather than its total loss.

# Methods

## Oblique-angle transmittance measurements

The oblique-angle transmittance measurements were carried out with an Agilent Cary 5000 spectrophotometer. The optical setup, which was assembled on a breadboard and placed into the spectrophotometer's sample compartment, is schematically shown in Fig. 2a. A set of aligned 0.8-mm pinholes (PH) separated by a 90-mm distance allows for obtaining a highly collimated beam (~0.5° angular width). Between the pinholes, a linear polarizer (LP) is mounted on a rotation stage to control the polarization of the incident beam. A wide aperture (A) and a shutter (SH), placed in front of the light source (LS) and detector (D), block stray light. The multilayer sample (S) is mounted on another rotation stage which controls the incidence angle of the input beam with a 0.5° precision. A 100% transmittance baseline spectrum was taken around 532 nm with a 0.04-nm bandwidth, 0.04-nm spectral step, and 5-s averaging time. A 0% transmittance reference spectrum was obtained after the scan with the shutter closed.

## Optical limiting measurements

The experimental apparatus for optical limiting measurements is shown in the schematics in Fig. 3a. The input beam is provided by a frequency-doubled Q-switched Nd:YAG laser and consists of 532-nm pulses of a temporal duration of ~6 ns. Each pulse is triggered individually (one shot every 30 s at low fluence and every 60 s at high fluence) to prevent possible sample heating by repetitive shots. The pulse energy is monitored by sending a weak reflected signal from a semi-transparent plate to a photodiode to obtain a reference for each laser pulse and proper normalization for transmitted and reflected pulse energies to mitigate the shot-to-shot energy fluctuations. The size of the laser beam is reduced to 1 mm by a diaphragm, which provides almost flat-top beam irradiation of the sample. The sample is mounted on a suitable rotary translation stage to select the irradiation area and align it to the input beam to optimize the transmittance and, at the same time, collect the reflected signal. The transmitted and reflected pulses are injected into different legs of a Y-fiber, sending both signals to the same photodiode. Because the fiber legs are of different lengths, resulting in a 125 ns delay between the two paths, both signals appear on the same trace of a digital oscilloscope connected to a photodiode. A series of filters in front of the sample allows setting the incident pulse energy or fluence to the desired value. When increasing the pulse energy, one or more filters ahead of the sample are removed and placed behind the sample. This allows the photodiode to always operate within an energy range that guarantees linearity, avoiding spurious, nonlinear responses. The filter sequence is chosen to cover more than four orders of fluence, from 25 μJ/cm$^2$ to 0.7 J/cm$^2$. The transmitted pulse energy, $T$, is normalized by the energy received in the absence of the sample, while the reflected pulse energy, $R$, is normalized by the energy collected when a mirror replaces the sample.

**Laser pulse propagation model and simulation**

The electromagnetic wave propagation in a medium stratified in the $z$-direction and having a thermo-optic effect is described by the following set of coupled electromagnetic wave and thermal equations:

$$\nabla^2 \boldsymbol{E} = \varepsilon\mu \frac{\partial^2 \boldsymbol{E}}{\partial t^2} + \mu\sigma \frac{\partial \boldsymbol{E}}{\partial t}, \quad \text{and} \quad \rho c \frac{\partial \theta}{\partial t} = \nabla \cdot (\kappa \nabla \theta) + \dot{Q},$$

where $\varepsilon = \varepsilon_r \varepsilon_0$, $\mu = \mu_r \mu_0$, $\varepsilon_0$ and $\mu_0$ are the permittivity and permeability of free space, respectively, $\varepsilon_r(z,\theta)$ and $\mu_r$ are the relative permittivity and permeability, $\sigma(z)$ is the electric conductivity, $c(z)$ is the specific heat capacity, $\rho(z)$ is the mass density, $\kappa(z)$ is the thermal conductivity, $\theta(z)$ is the temperature, and $\dot{Q} = \frac{1}{2}\sigma|\boldsymbol{E}|^2$ is the volumetric heat production rate.

In the modeling, we consider an incident linearly polarized plane wave in the form of a Gaussian pulse in the setting of Fig. 1a,

$$\boldsymbol{E}(z,t) = \boldsymbol{E}_0 \exp\left(-\frac{t^2}{2\sigma_t^2}\right) \exp\left[-\frac{\alpha z}{2}\right] \exp[i(\omega t - 2\pi n z/\lambda)], \qquad \boldsymbol{E}_0 \perp \boldsymbol{z},$$

where $\boldsymbol{E}_0$ is the peak field amplitude, $\sigma_t = 5$ ns is the pulse duration, $n$ is the refractive index, $\alpha = 4\pi k/\lambda$ is the absorption coefficient, and $k$ is the extinction coefficient, so that $n^2 - k^2 = \varepsilon_r \mu_r$ and $2nk = \mu_r \sigma/\omega\varepsilon_0$. We account for heat transport only along the wave propagation direction, which is equivalent to imposing periodic boundary conditions on the sides of the multilayer. At the same time, we impose convective boundary conditions on the front and back faces of the multilayer, assuming that the heat flux $q$ is dissipated as $q = h(\theta - \theta_0)$, where $h = 100$ W/(m$^2$K) is the convection heat transfer coefficient, and $\theta_0 = 296$ K is the ambient temperature. We use the following material properties: $\mu_r = 1$, $n_{r\,ZnS} = 2.39 + (dn/d\theta)(\theta - \theta_0)$[47], $k_{ZnS} = 2 \times 10^{-4}$[47], $\frac{dn}{d\theta} = 5.4 \times 10^{-5}$[48], $n_{r\,Cryolite} = 1.35$[47], $k_{Cryolite} = 0$[47], $c_{ZnS} = 0.527$ J/gK[48], $c_{Cryolite} = 1.04$ J/gK[49], $\kappa_{ZnS} = 27.2$ W/mK[48], $\kappa_{Cryolite} = 2.25$ W/mK[50].

The electromagnetic wave and thermal equations were solved numerically, using the coupled Microwave and Heat Transfer modules of a finite-element software package from COMSOL MULTIPHYSICS[51]. The convergence of the results has been evaluated with a tolerance factor of 0.1%. We then repeated the calculations by doubling the number of mesh points to guarantee the accuracy of the converged numerical solutions.

**Acknowledgements:** FR, DW, AB, FT, and SC thank Ente Cassa di Risparmio for financial support and Lorenzo Fini and Renato Torre for lab equipment and helpful suggestions. AB thanks the European Union - PON Research and Innovation 2014-2020 (MD.1062, August 10, 2021). AAC acknowledges the hospitality of the European Laboratory for Nonlinear Spectroscopy during his sabbatical visit and financial support from AFOSR (FA9550-19-1-0359, FA9550-20-F-0005). TK, RK, and SS acknowledge financial support from ONR (N00014-191-2480), NSF-RINGS ECCS (#2148318), and Simons foundation (MPS-733698). IV acknowledges financial support from AFOSR (LRIR 21RYCOR019).

## Supplementary Materials

### 1. Coupled-mode-theory of the planar three-mirror resonator.

A three-mirror resonator can be modeled as a system of two resonant modes $\omega_1$ and $\omega_2$, coupled to each other at rate $\kappa$ and to respective transmission lines at rates $\gamma_1$ and $\gamma_2$ (see Fig. 1a). The $2 \times 2$ scattering matrix $\boldsymbol{S}$ of this setting is given by

$$\boldsymbol{S}(\omega) = -\mathbf{1} - i\boldsymbol{W}\boldsymbol{G}\,\boldsymbol{W}^{\dagger}, \qquad \boldsymbol{G}(\omega) = (\boldsymbol{H} - \omega\mathbf{1})^{-1}, \qquad \text{(S1)}$$

where $\mathbf{1}$ is the $2 \times 2$ identity matrix, $\omega$ is the frequency of a monochromatic incident wave, $\boldsymbol{G}(\omega)$ is the Green's function, and $\boldsymbol{H} = \boldsymbol{H}_0 - \frac{i}{2}\boldsymbol{W}^{\dagger}\boldsymbol{W}$ is the non-Hermitian Hamiltonian consisting of two parts: $\boldsymbol{H}_0$ describing the isolated system and $\boldsymbol{W}^{\dagger}\boldsymbol{W}$ describing coupling to transmission lines via the coupling matrix $\boldsymbol{W}$. For the setting of Fig. 1a, we have

$$\boldsymbol{H}_0 = \begin{pmatrix} \omega_1 & \kappa \\ \kappa & \omega_2 \end{pmatrix} \text{ and } \boldsymbol{W} = \begin{pmatrix} \sqrt{2\gamma_1} & 0 \\ 0 & \sqrt{2\gamma_2} \end{pmatrix}. \qquad \text{(S2)}$$

We consider a scattering scenario in which an incident wave is injected into the system from the left. The reflection and transmission coefficients, $r(\omega)$ and $t(\omega)$, can be extracted from Eqs. (S1) and (S2) by applying the filtering matrices $\boldsymbol{P}_{in} = (1,0)$ and $\boldsymbol{P}_{out} = (0,1)$ to the $\boldsymbol{S}$-matrix. We have

$$r(\omega) \equiv \boldsymbol{S}_{11} = \boldsymbol{P}_{in}\,\boldsymbol{S}\boldsymbol{P}_{in}^{\dagger} \text{ and } t(\omega) \equiv \boldsymbol{S}_{21} = \boldsymbol{P}_{out}\,\boldsymbol{S}\boldsymbol{P}_{in}^{\dagger}. \qquad \text{(S3)}$$

We proceed by further analyzing $r(\omega)$. The (complex) frequencies $\omega_{RZ}$'s, for which $r(\omega_{RZ})$ becomes zero, correspond to specific solutions of the scattering problem for which there is no back reflection into the input channel, referred to as R-zeros (RZ). When $\omega_{RZ}$'s are real-valued, the corresponding scattering solutions describe monochromatic incident waves. Such solutions coincide with the reflectionless scattering modes (RMs) recently proposed[33,34] and implemented[52] in the framework of perfect-impedance-matching wavefront-shaping schemes. In the case of a passive system (no loss or gain), they can also be called perfectly transmissive modes, i.e., monochromatic incident waves with unity transmittance, $T(\omega_{RM}) \equiv |t(\omega_{RM})|^2 = 1$. We can show that the condition for the existence of such modes is expressed as[53]

$$r(\omega_{RZ}) = \frac{\det\left(\omega_{RZ}\mathbf{1} - \left\{\boldsymbol{H}_0 + i\left[W^{\dagger}P_{in}^{\dagger}P_{in}W - W^{\dagger}P_{out}^{\dagger}P_{out}W\right]\right\}\right)}{\det\left(\omega_{RZ}\mathbf{1} - \left\{\boldsymbol{H}_0 - i\left[W^{\dagger}P_{in}^{\dagger}P_{in}W + W^{\dagger}P_{out}^{\dagger}P_{out}W\right]\right\}\right)} = 0 \text{ and } \omega_{RZ} \in \mathbb{R}, \qquad \text{(S4)}$$

indicating that the requirement for the existence of RMs collapses to an eigenvalue problem for the auxiliary wave operator

$$\boldsymbol{A} = \boldsymbol{H}_0 + i\left[W^{\dagger}P_{in}^{\dagger}P_{in}W - W^{\dagger}P_{out}^{\dagger}P_{out}W\right] = \begin{pmatrix} \omega_1 & \kappa \\ \kappa & \omega_2 \end{pmatrix} + i\begin{pmatrix} \gamma_1 & 0 \\ 0 & -\gamma_2 \end{pmatrix}. \qquad \text{(S5)}$$

It is important to point out that the (complex) frequencies $\omega_{RZ}$'s that lead to zero reflection differ from the resonant frequencies $\omega_p$'s . The frequencies $\omega_p$'s are identified with the poles of the $\boldsymbol{S}$-matrix and are the complex roots of the secular equation, $\det(\omega_p\mathbf{1} - \boldsymbol{H}) = 0$, where $\boldsymbol{H} = \boldsymbol{H}_0 - i\left[W^{\dagger}P_{in}^{\dagger}P_{in}W + W^{\dagger}P_{out}^{\dagger}P_{out}W\right]$ appears in the denominator of Eq. (S4).

Since the wave operator $\boldsymbol{A}$ is non-Hermitian, its eigenvalues $\omega_{RZ}$'s are generally complex-valued. On the other hand, the RMs are a subset of purely real $\omega_{RZ}$'s. One way to enforce the reality of $\omega_{RZ}$'s is to impose dynamic symmetries on $\boldsymbol{A}$. Such a strategy has recently (on various occasions) been successfully implemented in the framework of the Hamiltonian $\boldsymbol{H}$ describing the resonant modes. Among the dynamic symmetries, the PT symmetry, where the parity operator $\boldsymbol{\mathcal{P}}$ represents a reflection with respect to a center of symmetry and the time-reversal operator $\boldsymbol{\mathcal{T}}$ represents the interchange of gain and loss within a system, was the tip of the spear in that newly designed schemes[28-32]. Here instead, we impose constraints on $\boldsymbol{A}$ to become PT-invariant, i.e., $[\boldsymbol{A}, \boldsymbol{\mathcal{PT}}] = 0$. This condition is satisfied provided (a) the coupling to the left and right transmission lines is the same, i.e., $\gamma_1 = \gamma_2 = \gamma$, and (b) the cavity resonance frequencies are equal to each other, i.e., $\omega_1 = \omega_2 = \omega_0$. The intuition provided by Eq. (S5) under these conditions helps to understand the PT nature of the auxiliary problem: the input channel acts as an effective "radiative gain" to the system, while the output channel functions as an effective "radiative loss." Indeed, the direct diagonalization of $\boldsymbol{A}$ under these conditions leads to the following results for the eigenfrequencies and eigenvectors,

$$\omega_{RZ}^{(\pm)} = \omega_0 \pm \sqrt{\kappa^2 - \gamma^2} \qquad \text{and} \qquad \psi_{RZ}^{(\pm)} \propto \left( i \mp \sqrt{\left(\frac{\kappa}{\gamma}\right)^2 - 1}, \frac{\kappa}{\gamma} \right)^T . \tag{S6}$$

For $\kappa \geq \gamma$, the spectrum is real, $\omega_{RZ}^{(\pm)} \in \mathbb{R}$, while $\boldsymbol{A}$ and $\boldsymbol{\mathcal{PT}}$ share the same set of eigenvectors. This parameter range corresponds to the so-called *exact PT-symmetric phase*. It is in this phase that the RMs are formed, $\omega_{RM}^{(1,2)} = \omega_{RZ}^{(\pm)} \in \mathbb{R}$, leading to unity transmittance for incident monochromatic waves with frequencies $\omega = \omega_{RM}^{(1,2)}$ (Fig. 1b). The other limiting case, $\kappa < \gamma$, is associated with the so-called *spontaneously broken PT-symmetry phase,* in which the spectrum consists of the complex-conjugate pairs $\omega_{RZ}^{+} = (\omega_{RZ}^{-})^*$. The transition to this domain is characterized by *spontaneous PT-symmetry breaking*[28], identified by the fact that the eigenvectors of $\boldsymbol{A}$ cease to be eigenvectors of $\boldsymbol{\mathcal{PT}}$, even though $\boldsymbol{A}$ and $\boldsymbol{\mathcal{PT}}$ commute. At the spontaneous symmetry breaking point, $\kappa = \gamma$, both the eigenfrequencies and the corresponding eigenvectors coalesce, forming an EPD. In contrast, the Hamiltonian $\boldsymbol{H}$ does not satisfy any PT symmetry nor demonstrates any EPD in its resonance spectrum.

The violation of the PT-symmetry turns the R-zeros $\omega_{RZ}^{(\pm)}$ from real to complex-valued, resulting in the destruction of the RMs with subsequent abrupt suppression of the transmittance (Fig. 1e,f) and significant enhancement of the reflectance. The symmetry violation can be spontaneous (as discussed above) or explicit. A physical mechanism that triggers *explicit PT-symmetry breaking* is based on self-induced detuning of one of the cavity resonance frequencies, $\omega_1 \to \omega_0 + \varepsilon$, due to, e.g., the Kerr or temperature-dependent nonlinearity that modifies the index of refraction of the cavity in the presence of high-power radiation. Another mechanism of PT-symmetry breaking is changing the angle of incidence, leading to a differential cavity frequency blue shift.

## 2. Cryolite/ZnS multilayer fabrication and characterization.

Figure S1a presents a scanning electron microscope (SEM) image of the cryolite/ZnS multilayer, where cryolite and ZnS films appear dark and grey, respectively. The SEM image reveals variation in film thicknesses due to an error compensation process in which the multilayer is continually optimized to compensate for "errors" made during the deposition. The ramification of this is a slight shift of $\omega_{RZ}^{(1,2)}$ from the real axis to the complex plane. Indeed, since the complex frequencies $\omega_{RZ}^{(1,2)}$ are zeroes of the complex field $r(\omega)$, they can be regarded as topological defects in the complex-frequency plane, which cannot disappear due to small perturbations of the photonic structure[54]. Thus, the complex shift of $\omega_{RZ}^{(1,2)}$, together with the optical loss in the multilayer, only contributes to the reduction in the peak transmittance without substantially affecting the passband formation. A normal-incidence transmittance spectrum obtained with an Agilent Cary 5000 spectrophotometer (Methods section) is shown in Fig. S1b by black-filled circles. The peak transmittance is reduced to 0.75 because of the glass substrates, structural imperfections, and photon absorption. The red dotted line represents the transfer-matrix-method[55] simulation transmittance for the model multilayer.

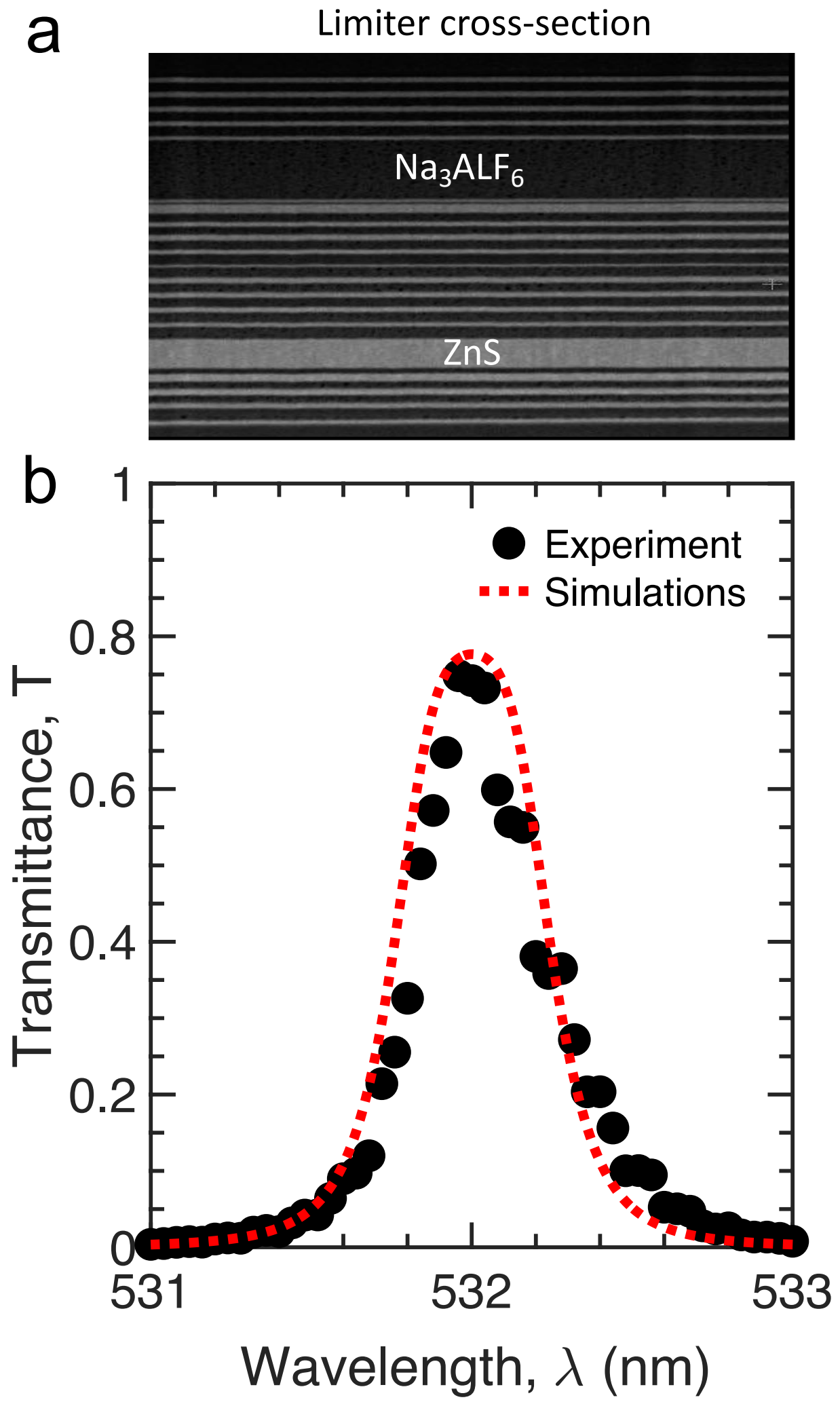

**Fig. S1: Cryolite/ZnS multilayer characterization. a,** SEM image of the cryolite/ZnS (dark/grey) multilayer. **b,** Normal-incidence transmittance measured with a spectrophotometer (black-filled circles) and obtained with transfer-matrix-method simulations (red dotted line). A quartic flat-top lineshape indicates the RM EPD.